\documentstyle[preprint,graphics,dcolumn,prd,aps]{revtex}
\begin{document}
\tightenlines
\draft
\title{Exponential Potentials for Tracker Fields}
\author{Claudio Rubano}
\address{Dipartimento di Scienze
Fisiche, Universit\`{a} di Napoli, Complesso Universitario di Monte S. Angelo,
Via Cintia, Ed. N, I-80126 Napoli, Italy
\\ and Istituto Nazionale di Fisica Nucleare, Sez. Napoli, Complesso Universitario
di Monte S. Angelo, Via Cintia, Ed.  G, I-80126 Napoli, Italy}
\author{Paolo Scudellaro}
\address{Dipartimento di Scienze
Fisiche, Universit\`{a} di Napoli, Complesso Universitario di Monte S.
Angelo, Via Cintia, Ed. N, I-80126 Napoli, Italy \\ and Istituto
Nazionale di Fisica Nucleare, Sez. Napoli, Complesso Universitario di
Monte S. Angelo, Via Cintia, Ed. G, I-80126 Napoli, Italy}
\author{Ester Piedipalumbo}
\address{Dipartimento di Scienze Fisiche,
Universit\`{a} di Napoli, Complesso Universitario di Monte S. Angelo, Via
Cintia, Ed. N, I-80126, Napoli, Italy \\ and Istituto Nazionale di
Fisica Nucleare, Sez. Napoli, Complesso Universitario di Monte S.
Angelo, Via Cintia, Ed. G, I-80126 Napoli, Italy}
\author{Salvatore Capozziello}
\address{Dipartimento di Fisica ``E. R. Caianiello'', Universit\`{a} di Salerno, 84081
Baronissi (Sa), Italy\\ and Istituto Nazionale di Fisica Nucleare,
Sez. Napoli, Complesso Universitario di Monte S. Angelo, Via Cintia,
Ed. G, I-80126 Napoli, Italy}
\author{Monica Capone}
\address{Dipartimento di Scienze Fisiche,
Universit\`{a} di Napoli, Complesso Universitario di Monte S. Angelo, Via
Cintia, Ed. N, I-80126, Napoli, Italy}
\date{\today}
\maketitle
\begin{abstract}
We show that a general, exact cosmological solution, where dynamics of
scalar field is assigned by an exponential potential, fulfils all the
issues of dark energy approach, both from a theoretical point of view
and in comparison with available observational data. Moreover,
tracking conditions are discussed, with a new treatment of the well
known condition $\Gamma>1$. We prove that the currently used
expression for $\Gamma$ is wrong.
\end{abstract}
\pacs{PACS number(s): 98.80.Jk, 98.80.Cq, 98.80.Hw, 04.20.Jb}
\narrowtext

\section{Introduction}

\subsection{Motivations}

The recent discovery of the acceleration of the Universe \cite{1,2}
has produced an overwhelming number of papers on its possible
explanations, alternative to a plain cosmological constant (see Refs.
\cite{3,4,4bis} for reviews). Historically, the most favourite one is
the introduction of a scalar field, usually minimally coupled with
gravity (but see Refs. \cite{4bis1,4bis2,5,6,7,8,9,9bis} for other
possible models). This poses the problem of the choice of a suitable
potential. Moreover, as the equations are not exactly solvable in
general, there is also a problem of qualitative analysis of the
solutions and of the choice of initial conditions. Usually, very crude
approximations are made, first of all the socalled \textquotedblleft
slow roll\textquotedblright\ condition, i.e., $\dot{\varphi}^{2}\ll
V(\varphi )$.

The aim of this paper is to illustrate some subtleties involved by this
situation, by means of a complete treatment of the general exact solutions
of the Einstein equations in presence of both nonrelativistic matter and
scalar field, with a suitable physical potential, and with a new treatment
of the well known \textquotedblleft tracking condition\textquotedblright, $%
\Gamma>1$ .

One result is that the slow roll condition is indeed not necessary and that
insisting on it can bring to lose very interesting models; closely connected
to this, is a discussion on the meaning of the statement that the state
equation $w\equiv p_{\varphi }/\rho _{\varphi }$ for the scalar field should
be ``almost constant''. By means of an analytic example, we shall see that
this expression is highly misleading, indeed.

Our choice for the potential is a particular class of exponential functions
\begin{equation}
V(\varphi)=B^{2}e^{-\sigma\varphi} \,,  \label{eq1}
\end{equation}
with $B$ arbitrary and $\sigma\equiv\sqrt{3/2}$ (in units $8\pi G=1$). The
reason for this special choice will be clear below.

First of all, we have to say that, although exponential potential is
natural in higher dimension theories, supergravity and superstring
models \cite{9ter,9quater,9quinque}, it is not generally considered as
feasible, even though it is often used at least as a first tool of
analysis \cite{10,12,11}. The usual objection to this potential is in
fact that it cannot really be the right one, because \textquotedblleft
it scales approximately\textquotedblright\ like the matter component,
so that, if the dark energy is dominant now, it should have been
dominating also in the past, which is in contrast with BBN
observations \cite{13}. This potential was nonetheless proposed by
some of us in Ref. \cite{14}, where also some explanations of the
reason why this is not true in our case are given. Anyway, in Sec. III
the problem is fully explored, and we show why the above objection is
wrong.

A second common objection is that it describes a very peculiar
situation, due to a fine tuning of initial conditions. We insist on
the point that the solution found in Ref. \cite{14} is general and
exact. Here, we start indeed from a particular choice, but then we go
further and release it, showing that the same \textquotedblleft final
situation\textquotedblright\ is reached, starting from a wide set of
\textquotedblleft initial conditions\textquotedblright. Again, we use
quotation marks because the expressions are ambiguous and misleading,
so that a preliminary discussion is needed to clarify what we mean.

A third possible objection is that this model could be in contrast
with the observations. As a matter of fact, some papers have already
appeared, successfully comparing this model with present day
observations \cite{15,16,17}. In Sec. II.B a first comparison with
observational data is made, with very encouraging results. However,
this paper is mainly concentrated on theoretical discussion. Our
purpose is to show that the model we consider can emulate very well a
standard model with dust plus a cosmological constant, well beyond any
reasonable improvement of the observational techniques. In order to do
this, a preliminary discussion on what is really measurable is again
needed.

A second important result of our considerations consists in showing that
many of the current statements about tracking solutions are based on an
incorrect result and should therefore be revisited.

The paper is organized as follows. First, we examine some issues
usually investigated when building a cosmological model with dark
energy; key features like cosmic components, choice of the scalar
field potential, measurability, and tracking behavior are therefore
shortly sketched. Sec. II is then devoted to the exponential potential
case and the general, exact solution that can be given to the
cosmological equations. A comparison with a constant $\Lambda $ model
and observational data is made; also, there can be found an analysis
of the scalar field equation of state and a more detailed one of the
tracking behavior. In Sec. III, the features usually assigned to the
tracking notion are critically examined, giving new insights into such
a concept. Finally, concluding remarks are drawn in Sec. IV.

\subsection{Preliminary discussion}

First of all, before going through the analysis of our potential, we
think it is useful to discuss some general questions which should be
of help in understanding better what follows.

\subsubsection{How can we set the choice for the potential?}

At the moment there is no idea of which form the potential of the
scalar field should take. The widely discussed inverse power law
\cite{4bis,20}, for instance, is certainly useful for a qualitative
treatment of the problem, but only under the assumption of the slow
rolling condition. It seems that the answer to this problem is simple
in principle: the right form is the one that gives the best fit to the
observed luminosity distance curve $d_{L}(z)$, and the fact that we
cannot establish it firmly is only due to our limited observational
precision. Indeed, it is possible to show that, whichever is the
expression of $d_{L}(z)$, we can find, in principle, a suitable form
for $V(\varphi)$, which gives it back. Unfortunately, the value of
$\Omega_{0m}$ turns out to be \textit{arbitrary}! The reason of this
is due to the fact that the introduction of an arbitrary function in
the model is equivalent to taking an infinite number of free
parameters into account. Worst of all, this result is possible only if
one has the \textit{exact} functional form of $d_{L}(z)$. This is
clearly impossible without already having a model in mind, so that one
is led again at the starting blocks \cite{20bis1,20bis2,20bis3,21,22}.

In our opinion, the only possible way to escape from this \textit{impasse}
is to ask for potentials which have some possible explanation in terms of
fundamental physics or which exhibit some nice mathematical feature, which
at least leaves the hope that it is due to some still unknown fundamental
physical property. As for the first requirement, to our knowledge, the best
candidates are the exponential potential, like that in Eq. (\ref{eq1}) (but
with $\sigma$ undetermined), and polynomials of the form
\begin{equation}
V(\varphi)=\frac{1}{2}m^{2}\varphi^{2}+\lambda\varphi^{4} +\text{\textit{%
hig. ord.}}  \label{eq2}
\end{equation}

It is important to observe, however, that this is a generic expansion of an
even function $f(\varphi)$, with a suitable minimum, so that any function of
this type will be good as well.

As for the second requirement, we shall see below that the potential
proposed in Eq. (\ref{eq1}), due to the special value of $\sigma$, exhibits
a Noether symmetry in the equations, which is just the reason why exact
integration is possible. Also, the potential of Eq. (\ref{eq2}) can be seen
as the expansion of a generalized hyperbolic sine, with the same type of
symmetry \cite{23} (see below).

On the other hand, it is also possible to show that a sum of
exponentials of the class in Eq. (\ref{eq1}) plus a cosmological
constant is essentially the most general type which can lead to exact
integration in elementary way \cite{24}.

\subsubsection{What is really measurable?}

In the context chosen by us, we have a two-dimensional configuration space,
with variables $\left\{ a,\varphi \right\} $ plus the \textit{velocities} $%
\left\{ \dot{a},\dot{\varphi}\right\} $, so that the total number of
variables is four. Consequently, this is the maximum number of independent
quantities which can be measured. As a matter of fact, none of these
quantities is in fact practically measurable. In flat cosmologies, $a_{0}$
is not measurable \textit{in principle}. This is connected with the choice $%
k=0$, i.e., in terms of measurable quantities, $\Omega _{tot.}=1$ exactly,
which is clearly artificial. Instead of these rather abstract objects, we
can also use derived quantities. One of them is, of course, $\Omega _{0m}$.
This parameter is very important, not only for its physical meaning, but
also because its value can, in principle, be measured independently from the
SNIa observations \cite{25}. Much has also been said about the value of the
acceleration parameter $q_{0}$; but, again, it is not directly measurable in
practice. However, $q_{0}$ is also present in the second term of the series
\begin{equation}
d_{L}(z)=H_{0}^{-1}\left( z-\frac{(1-q_{0})}{2}z^{2}+...\right) \text{ }\,,
\label{eq3}
\end{equation}%
so that it seems possible to measure it, after all. Unfortunately,
truncation of Eq. (\ref{eq3}) after the second term is valid only at rather
low redshifts (say $z\leq 0.1$). At these values, it is practically
impossible to obtain sufficient precision, while at higher values the other
terms cannot be neglected.

What is thus left? It seems reasonable to stick to what is for the
moment firmly established, i.e., $h\approx0.7$ and
$\Omega_{0m}\approx0.3$, with $3\sigma$ level $\sim\!10\%$ and
$\sim50\%$ errors, respectively\footnote{These values are based on
estimates not strictly dependent on the cosmological model.}. If we
are very optimistic, we can expect an improvement of precision in the
future of a factor 5, say. We shall see that even this could be not
enough to discriminate among models.

Also, a third useful variable could be the age of the universe
$t_{0}$, for which we have only a rather poor estimate $t_{0}>12$ Gyr.

\subsubsection{Which is the precise meaning of the expression
\textquotedblleft tracking behavior\textquotedblright ?}

One of the greatest advantages of the introduction of a scalar field
consists in the possibility to avoid the fine tuning of initial conditions,
obtaining the same final behavior \textquotedblleft for a wide range of
initial possibilities\textquotedblright. The last sentence is intentionally
rather ambiguous. In fact, the situation is often depicted as
\textquotedblleft similar but not exactly equal\textquotedblright\ to that
of an attractor in the theory of dynamical systems.

Let us thus try to be more precise. The equations which we have to
consider are the Einstein equations plus the Klein-Gordon equation for
the scalar field (for sake of simplicity we write them directly in the
flat case, but the following argument holds in the general case as
well)
\begin{equation}
3H^{2}=\rho _{\varphi }+\rho _{m}\text{ } ,  \label{eq4}
\end{equation}

\begin{equation}
2\frac{\ddot{a}}{a}+\left( \frac{\dot{a}}{a}\right) ^{2}+\frac{1}{2}\dot{
\varphi}^{2}-V(\varphi )=0\text{ } \,,  \label{eq5}
\end{equation}

\begin{equation}
\ddot{\varphi}+3H\dot{\varphi}+V^{\prime }(\varphi )=0\text{ } \,.
\label{eq6}
\end{equation}

Eq. (\ref{eq4}) is a first integral of the other two. The integration
constant is hidden into $\rho _{m}$. For instance, in the case of dust, we
have $\rho _{m}=Da^{-3}$; again, the value of $D$ is not directly
measurable, but we shall not in fact make use of this equation. Instead, we
shall use Eqs. (\ref{eq5}) and (\ref{eq6}). The reason is that it can be
shown that, when $\rho _{m}$ is related to a pure dust component, they are
derived by the following Lagrangian \cite{26,27}
\begin{equation}
L=3a\dot{a}^{2}-\frac{1}{2}a^{3} \dot{\varphi}^{2} +a^{3}V(\varphi )\text{}
\,.  \label{eq7}
\end{equation}
Eq. (\ref{eq4}) is then obtained as conservation of the energy function $%
E_{L}\equiv \left( \partial L/\partial \dot{a}\right) \dot{a}+\left(
\partial L/\partial \dot{\varphi}\right) \dot{\varphi}-L$.

A first consequence of this fact is that Liouville's theorem holds for this
system. Thus, the phase space volume of possible initial conditions is
conserved and has no attractors.

Nevertheless, the tracking behavior can be recovered by the following
argument. Consider a set of independent variables for this system (not
necessarily $\{a,\varphi ,\dot{a},\dot{\varphi}\}$) and a volume of possible
initial conditions. During the evolution, the volume is deformed, and it is
possible that its projection on a two(three)-dimensional subspace converges
to a point or a line. It is clear now that the result is strongly dependent
on the choices of variables and projection. It is also clear that the gain
in information on the variables of the subspace is compensated by a loss of
information on the variables of the complement.

In our specific case we can take advantage of the fact that, as said above,
only a part of the possible variables is practically measurable. Therefore
we can make the usual choice for the subspace, i.e., consider $\{\log a,\log
\rho _{\varphi }\}$, and show that, for a large set of initial conditions of
these two variables\footnote{%
Apparently, $a$ is constrained to be zero at the initial point, but in fact
the models cannot be pushed up to the initial singularity, so that the
initial time cannot be really zero. This gives some arbitrariness in the
choice of initial $a$. (See next Section for details.)}, the orbits converge
to a line. Another possibility is to consider $\{H,\Omega _{m}\}$, and check
that the orbits converge to a point.

It is also clear that only \textit{observable} quantities should be taken
into account. For any other parameter the concept of fine tuning is, in our
opinion, simply meaningless.

The situation is in fact rather involved and will be discussed in Sec. II.D,
by means of an example.

We think that the concept illustrated above is the only one which is
relevant for the solution of the coincidence problem. However, in the
literature the term \textquotedblleft tracking\textquotedblright\ is often
used in a different meaning. The idea is that a large set of initial
conditions should converge towards a \textit{nearly constant} $w$, well
\textit{before} the present epoch, and maintain this condition forever. We
shall show in Sec. II.C that this situation, although of course possible, is
not at all necessary.

\section{The exponential potential case}

Let us here introduce the exponential potential in the cosmological
framework. As we shall see, in this case the equations are easily
integrable, which allows many interesting considerations. (See Ref.
\cite{25bis} for a more general treatment). After finding the
solution, we derive some consequences with respect to observational
data and special features of quintessence field.

\subsection{The solution}

As said above we start from Eqs. (\ref{eq5}) and (\ref{eq6}), with potential
(\ref{eq1}), and, due to the existence of a Noether symmetry, we choose to
make the transformation of variables
\begin{equation}
a^{3}=uv\quad ,\quad \varphi =-\frac{1}{\sigma }\log \frac{u}{v}\text{ } \,.
\label{eq8}
\end{equation}

With these new variables, the Lagrangian (\ref{eq7}) becomes
\begin{equation}
L=\frac{4}{3}\dot{u}\dot{v}+B^{2}u^{2}\text{ } \,,  \label{eq9}
\end{equation}
and we see that the variable $v$ is cyclic, showing the existence of a
Noether symmetry. The related conserved quantity is
\begin{equation}
F\equiv \frac{\partial L}{\partial \dot{v}}=\frac{4}{3}\dot{u}=\frac{1}{3}(6
\dot{a}+\sqrt{6}a\dot{\varphi})(a\exp (-\sigma \varphi ))^{1/2}
\end{equation}
and makes the integration of the equations immediate, so that we get
\cite{14}
\begin{equation}
u=u_{1}t+u_{2}\text{ } \,,  \label{eq10}
\end{equation}

\begin{equation}
v=\frac{1}{6}u_{1}\omega t^{3}+\frac{1}{2}u_{2}\omega t^{2}+v_{1}t+v_{2}
\text{ } \,,  \label{eq11}
\end{equation}
with $\omega \equiv \sigma ^{2}B^{2}$. We thus have a general exact solution
with four integration constants $u_{1}$, $u_{2}$, $v_{1}$, $v_{2}$. It is
not easy to treat this solution in full generality, and, as we have seen,
not very useful. We have, therefore, to make a clever choice.

First, we fix the origin of time by setting $a(0)=0$. This can be done
in full generality. It is in fact important to observe that we are not
really fixing an initial condition for $a$, since the model is not
valid at initial singularity. The initial value of $a$ is fixed by the
time $t_{in},$ which we choose for the beginning of validity of the
model itself, and which is to some extent arbitrary. (See Sec. II.D
for a discussion on this point.) Thus, we have to set $u_{2}=0$,
\textit{or} $v_{2}=0$. It is easy to convince oneself that it is
arbitrary to decide which one. But in Ref. \cite{14} we made more than
this, and decided to set to zero
\textit{both}. For the moment we make the same choice, i.e.,
$u_{2}=v_{2}=0$. This in fact selects the tracking solution, but this
will be better examined in Sec. II.D, where such a condition will be
released.

The second condition which we set is $t_{0}=1$. That is, we fix the time
scale by taking as unit the age of the universe. Indeed, as $t_{in}$ is
unknown, this is not exactly the age of the universe, but the difference is
irrelevant. In this way we get rid of a very badly known quantity. It would
be possible, anyway, to avoid this condition, but we have checked that there
is no substantial advantage.

The third condition is to set $a_{0}\equiv a(t_{0})=a(1)=1$. This is
standard and fixes the normalization of $a$, according to the above
discussion.

The last condition is to set $H(t_{0}=1)\equiv {\mathcal{H}}_{0}$. Due
to the choice of $t_{0}$, this parameter is of order one, \textit{but
it is not the same as the usual} $h$. To be clearer, since we are
using a more or less arbitrary unit of time, this parameter gives no
information on the observed value for $H_{0}$, but, as we shall see,
it actually fixes the value of $%
\Omega _{0m}$. An independent measure of $H_{0}$, together with the value of
${\mathcal{H}}_{0}$, gives the age of the universe for this model.

The relevant physical quantities are then
\begin{equation}
H=\frac{\left( 12{\mathcal{H}}_{0}-8\right)
t^{2}+8-6{\mathcal{H}}_{0}}{\left( 9 {\mathcal{H}}_{0}-6\right)
t^{3}+\left( 12-9{\mathcal{H}}_{0}\right) t}\text{ }
\,,  \label{eq14}
\end{equation}

\begin{equation}
\Omega _{m} =\frac{\left( 4-3{\mathcal{H}}_{0}\right) \left( \left( 3{\mathcal{
\ H}} _{0}-2\right) t^{2}-\left( 3{\mathcal{H}}_{0}-4\right) \right) }{\left(
2\left( 3 {\mathcal{H}}_{0}-2\right) t^{2}-\left(
3{\mathcal{H}}_{0}-4\right)
\right) ^{2}}\text{ } \,,  \label{eq15}
\end{equation}

\begin{equation}
w=-\frac{2\left( 3{\mathcal{H}}_{0}-2\right) t^{2}+3\left(
4-3{\mathcal{H}}
_{0}\right) }{4\left( 3{\mathcal{H}}_{0}-2\right) t^{2}+3\left( 4-3{\mathcal{H}}
_{0}\right) }\text{ } \,.  \label{eq16}
\end{equation}

It is possible to see that the constant $u_{1}$ is present only in the
expression of $\varphi _{0}$
\begin{equation}
\varphi _{0}=-\sqrt{\frac{2}{3}}\log u_{1}^{2}\,.  \label{eq17}
\end{equation}
All the other quantities only depend on ${\mathcal{H}}_{0}$, so that
it will also fix the values of $\Omega _{0m}$ and $w_{0}$ according to
the equations
\begin{equation}
\Omega _{0m}=-\frac{2\left( 3{\mathcal{H}}_{0}-4\right) }{9{\mathcal{H}}_{0}^{2}}
\quad ,\quad w_{0}=-\frac{8-3{\mathcal{H}}_{0}}{4+3{\mathcal{H}}_{0}}\,.
\label{eq18}
\end{equation}

For ${\mathcal{H}}_{0}=1$, we have $\Omega _{0m}\approx 0.22$ and $
w_{0}\approx -\,0.71$. These values are reasonable and in good
agreement with the ones found in Ref. \cite{15} for a best fit with
SNIa data; they also agree at $1\sigma $ level with the values we will
find below. Assuming $h=0.7$, we get $t_{0}\approx 15\,Gyr$, which is
also good. Moreover, we get
\begin{equation}
\frac{\dot{\varphi}_{0}^{2}}{2V(\varphi _{0})}=\frac{1}{6}\text{ } \,,
\end{equation}
showing that the slow rolling condition is not fulfilled, as said in the
introduction.

When one is not interested in precise fits, but instead on qualitative
analysis (like in Sec. II.D), it is useful to stick to these values,
so that all the formulae get simpler. We thus get
\begin{equation}
a^{3}=\frac{t^{2}\left( t^{2}+1\right) }{2}\quad ,\quad
H=\frac{4t^{2}+2}{ 3t^{3}+3t}\quad ,\quad
\Omega_{m}=\frac{t^{2}+1}{\left( 2t^{2}+1\right) ^{2} }\quad ,\quad
w=-\frac{2t^{2}+3}{4t^{2}+3}\,.
\end{equation}

\subsection{Comparison with $\Lambda$-term model and observations}

We want, now, to show that our model is practically equivalent to a
$\Lambda$-term model in accounting for present day observations. Since
the aim of this paper is mainly theoretical, we present here only a
preliminary check, postponing to a forthcoming paper a thorough
treatment of various popular models.

In order to do this we take advantage of the CAMB \cite{27a} and
CosmoMC \cite{27b} codes. The powerful CosmoMC algorithm allows to
compare a model with the Wilkinson Microwave Anisotropy Probe (WMAP)
\cite{27b1} and CBI \cite{27b2} data, as for the CMBR spectrum, and
with SNIa data and other constraints coming from BBN and HST Key
Project. The LSS power spectrum derived from the Two Degree Field
(2dF) Galaxy Redshift Survey \cite{27b3} is also taken into account.

In short the procedure consists in generating, from random sets of the
parameters, thousands samples of results. The maximum likelihood is
then obtained by product of the single likelihood functions as
established by the observers teams. The details of the treatment can
be found in Refs. \cite{27c,27d}.

It is important to observe that the results are obtained by numerical
integration of the equations, by means of a CAMB subroutine which
\textit{includes radiation. }This is necessary for generating CMBR
spectrum, starting from the radiation dominated era. On the other
hand, this should not affect the previous discussion, which in fact
refers to a period well after decoupling. In any case, a first
comparison of the effective $w$, ${\bar{w}}\equiv \int w\Omega
_{m}da/\int \Omega
_{m}da$, and of the function $w(a)$, obtained with numerical
(including radiation) and analytical (without radiation) integrations
of our model, gives perfectly overlapping results.

After running the program we obtain the best fit results reported in
Table I. It is clear that the two sets of parameters are practically
the same. Anyway, it is possible to compare the $\log(LH)$ values,
from which it appears that our model is slightly, but of course not
significantly, disfavoured.

Note that in our model it is possible to get an analytic parametric
expression for the luminosity distance in terms of hypergeometric and
Euler functions. We find

\begin{equation}
z=\left( \frac{2}{t^{2}\left( t^{2}+1\right) }\right) ^{1/3}-1 \,,
\label{eq20}
\end{equation}

\begin{equation}
d_{L}(t,{\mathcal{H}}_{0}=1)=\frac{1}{\left( t^{2}+1\right)
^{1/3}}\left(
\frac{3 \sqrt{\pi}\Gamma(7/6)}{\Gamma(2/3)t^{2/3}}-\frac{2^{2/3}
\,_{2}F_{1}(1/6,1/3,7/6,-t^{2})}{t}\right) \,.  \label{eq21}
\end{equation}

From this it is not difficult to compute the distance modulus and compare it
with $\Lambda$-term model and SnIa data, which is done in Fig. 1.

With the aid of CAMB code, we have also generated the CMBR spectrum for the
two models, which is presented in Fig. 2.

The overlapping of the plots is striking in both cases. It is
interesting to note that in this test we obtain equal values for
$\Omega _{m}$ and marginally different values for $h$. In a previous
test, based only on SnIa \cite{15}, we obtained instead very similar
values\footnote{This may be not so clear in Ref. \cite{15}, where the
value of $h$ is hidden in the parameter $m_{0}$, which in fact turns
out to be the same as Perlmutter's value.} for $h$  and quite
different values for $\Omega _{m}$.

It appears very unlikely that more precise observations, of the same
type here considered, will disentangle this degeneracy. An independent
precise measure of $h$ and/or $\Omega _{m}$ could be, on the contrary,
of great help.

\subsection{Analysis of $w$}

Let us now analyze the variation of $w$, in order to understand better what
was said in Sec. I.B.

First of all, let us observe that the very notion of \textit{almost constant}%
\ is somewhat ambiguous: we have in fact to ask \textit{almost constant with
respect to which independent variable}? We see from Fig. 3 that the
function $w\left( t\right) $ does not show any flatness except towards
infinity, which can be also easily deduced from Eq. (\ref{eq16}).
Quite different is the situation of $w\left( \log a\right) $, showed
in Fig. 4. Here we have $ w \sim -1$ in the past, with a sharp
transition to the asymptotical value $w=-0.5$. What is most intriguing
is the fact that the present time (i.e., $a_{0}=1$), marked by the
vertical line, occurs just in the middle of this transition. Thus, we
see that for the range of observational data, we have to consider the
epoch when $w$ is mostly varying, that is, just the contrary of what
is generally considered! The situation is illustrated also by the plot
in Fig. 5, which contains the same type of information in a perhaps
more familiar form.

Let us make some remarks on this point.

i) The situation just described looks like a striking cosmic
coincidence, but we can show that it is not so \textit{striking}.
Indeed, the plot in Fig. 4 is done with the best fit values of Table
I, but we can vary the present value of $\Omega _{0m}$ over a quite
large range, much wider than any imaginable one, and obtain very
similar results, as illustrated by Fig. 6.

ii) The above example shows that, \textit{in the context of our
model}, the allowed value of the parameter forces to place the present
epoch in the period of maximal variability of $w$, \textit{with
respect to} $\log a$. If we consider another dependence, and of course
if we change the model, this is not necessarily true.

iii) This example should indicate that the arguments involving variations of
$w$ are really very subtle, and that it may be dangerous to base our
conclusions on rough approximations based on this type of arguments.

\subsection{The tracking behavior}

In this section, we release the condition $u_{2}=v_{2}=0$, which gave
us the particular solution analyzed so far. As said above we can
safely set $a(0)=0$, i.e., one of the two parameters $u_{2}$ or
$v_{2}=0$. We have verified that it is better to set $u_{2}=0,$ since
we get simpler formulae, which are rather complex anyway. Thus, since
we are not interested here in best fit values, we set from now on
directly ${\mathcal{H}}_{0}=1$, in order to obtain simpler
expressions. As before, we also set in full generality $t_{0}=1$ and
$a_{0}=1$, so that we are left with $u_{1}$ (as before) and $v_{2}$.
Being it arbitrary, we pose $\varepsilon\equiv u_{1}v_{2}$, so that we
have
\begin{equation}
u=u_{1}t\quad,\quad v=\frac{1}{6}u_{1} \omega t^{3}+v_{1}t+\varepsilon/u_{1}
\,,
\end{equation}
and, after evaluating $\omega$ and $v_{1}$ according to our conventions, we
obtain for the relevant objects
\begin{equation}
a^{3} =\frac{1}{2}t^{2}(t^{2}+1)+\frac{1}{2}t(2-3t+t^{3})\varepsilon \,,
\end{equation}

\begin{equation}
\varphi =\sqrt{\frac{2}{3}}\log\left( \frac{t+t^{3}+(2-3t+t^{3})\varepsilon
} {2u_{1}^{2}t}\right) \,,
\end{equation}

\begin{equation}
H =\frac{2}{3t}\frac{t+2t^{3}+(1-3t+2t^{3})\varepsilon}{t+t^{3}
+(2-3t+t^{3})\varepsilon} \,,
\end{equation}

\begin{equation}
w=-\frac{2t^{6}+3t^{4}+(4t^{6}-6t^{4}+8t^{3})\varepsilon
+(2t^{6}-9t^{4}+8t^{3}-1)\varepsilon^{2}}{4t^{6}+3t^{4}+(8t^{6}
-6t^{4}+4t^{3})\varepsilon+(4t^{6}-9t^{4}+4t^{3}-1)\varepsilon^{2}} \,,
\end{equation}

\begin{equation}
\Omega_{m}=\frac{t(t^{3}+t)-t(2t^{3}+6t-2)\varepsilon-t(3t^{3}
-9t+6)\varepsilon^{2}}{4t^{6}+4t^{4}+t^{2}+(8t^{6}-8t^{4}+4t^{3}
-6t^{2}+2t)\varepsilon+(4t^{6}-12t^{4}+4t^{3}+9t^{2}-6t+1)\varepsilon^{2}}
\,.
\end{equation}

We see that, as before, $u_{1}$ alone only enters in the present value
of $\varphi$, which is again given by Eq. (\ref{eq17}). The observable
quantities are now parametrized by $\varepsilon$, which must be small,
if we want to recover the unperturbed solution discussed above. But it
is important to say that, in our opinion, a small value of
$\varepsilon$ should not be interpreted as a fine tuning on initial
conditions. In fact, $\varepsilon$ is just a mathematical parameter;
thus, according to what we said above, there is no point in discussing
fine tuning of it. In any case, as we shall see just below, its value
can span the range $0\div10^{-9}$, so that it actually covers many
orders of magnitude.

A second point must be discussed now. The presence of $\varepsilon
\neq 0$, no matter how small, dramatically changes the solution near
the singularity; we have indeed\footnote{For deeper considerations on
the stability of solutions of this type, see Ref. \cite{27e}.}
\begin{equation}
\lim_{t\rightarrow 0}w=1\quad ,\quad \lim_{t\rightarrow 0}\Omega _{m}=0\,,
\end{equation}
and, depending on the value of $\varepsilon $, these conditions can be
approximately maintained until a rather recent epoch. This is clearly
not acceptable, and forces us to take very small $\varepsilon $, which
sounds good. The problem is: how small? A reasonable answer seems to
push the region where $w\approx 1$ farther than $z\approx 10^{3}$,
i.e., before decoupling, where our model may be not good to work. We
may consider, then, three possible strategies.

1) To be very restrictive and say that, before decoupling, the model is
clearly not valid. In this way we shall see that we have no tracking
behavior at all, but there is general agreement that convergence towards the
tracking solution should begin \textit{before} decoupling, so that this
attitude does not bring us much farther.

2) To be less restrictive and say that the model can be anyway roughly good
after equivalence, and just before decoupling. This corresponds to set the
initial point somewhere between $z\approx10^{3}$ and $z\approx10^{4}$.

3) To be still less restrictive and say that, at least as toy model (i.e.,
as mathematical illustration of the situation), we can try to analyze what
happens until nearby the initial singularity.

In the following we are discussing these two last situations, trying to
clearly distinguish among them.

Let us start from Fig. 7, where we present the complete plot of $w$
versus $\log _{10}a$, for a particular very small value of
$\varepsilon $. We see that the transition between $w\approx 1$ and
$w\approx -1$ occurs at $z\approx 10^{3}$, as requested. Let us now
vary $\varepsilon $ in the range%
\footnote{
As said above, $\varepsilon $ can take values up to zero, but the
smaller values are difficult to plot and make the figure less clear.}
$10^{-12}\div 10^{-9}$, which gives the plot in Fig. 8 (only the first
transition is now shown). If we first adopt the point of view 2), we
see that an almost complete set of values $w\in (-1,1)$ evolves
towards $w=-1$ (and then the second transition), with irrelevant
differences. The situation does not change if we adopt strategy 3), in
which case however the Fig. 8 may be misleading. In fact, as
$\varepsilon $ gets smaller and smaller, all the region with $w<1 $ on
the left should be filled with lines.

Now we pass to the more familiar plot of $\log_{10}\rho_{\varphi}$
versus $\log_{10}a$ (Fig. 9). Again, in the case 2), we can set the
initial point between $z\approx10^{3}$ and $z\approx10^{4}$ and see
that the values of $
\log_{10}\rho_{\varphi}$ span the interval $(0,0.6)$, so that $
\rho_{\varphi} $ can vary upon about $6$ orders of magnitude. This is not
the hundreds of them presented in the literature, but we can recover this
possibility just by adopting strategy 3). In this case, we have to consider
all the triangle on the left until, say, the end of inflation.

In Fig. 10 we present the plot of $H$ versus $\Omega _{m}$. The
situation seems very similar to the others, with a lot of initial
conditions converging towards the tracker solution. But, if you look
at the values of $\Omega _{m}$ on the scale,  all the values are
indeed very close to $1$. Only if we adopt strategy 3) and go very far
in the past, we get a somewhat larger set of initial values for this
parameter. From a physical point of view this is good news, because we
want $\Omega_{m}$ to dominate during the radiation epoch. This
suggests that option 3) is not so bad, and indeed, as said before, a
first verification on the function $w(a)$, obtained as a byproduct of
CAMB integration including radiation, gives almost perfect concordance
up to a redshift $\sim 10^{6}$.

From a pedagogical point of view, together with the above cases, we learn
that the tracking behavior can be different, depending on which variables
are taken into consideration, so giving strength to our initial
consideration on the necessity to be very careful when treating this
argument.

Needless to say, the tracking behavior will be completely lost if we
consider the variable $\varphi$, and leave undetermined its present
value. If we decide to fix it, which in principle can always be done,
we have to set $u_{1}$ independently. If we set it to $1$ (but this is
relevant only for numerical computation of the plot) and look at Fig.
11, for instance, the situation is quite similar to that of
$\Omega_{m}$.

\section{The tracking concept revisited}

We are now ready to analyze some widely accepted concepts about tracking
solutions and understand some subtleties and incorrect results.

As said above, a currently accepted point of view on tracking
solutions is that the "attractor" should be characterized by a nearly
constant value of $ w $, well before the present epoch. Of course, it
is true that the tracking solution must be reached before, but the
above example should make it clear

i) that it does not imply $w$ being constant, while, on the contrary, the
present epoch occurs in the period when $w$ changes most;

ii) that the asymptotic constant value has not necessarily anything to do
with observations.

A second current point of view is that the potentials which are good for
tracking should be such that
\begin{equation}
\Gamma \equiv \frac{VV^{\prime \prime }}{(V^{\prime })^{2}}>1\,,
\end{equation}
with $\Gamma (\varphi )$ nearly constant over the possible range of $\varphi
$. In this case, after some oscillations between $-1$ and $1$, $w$ should
reach a stable nearly constant value, given by
\begin{equation}
w=\frac{w_{B}-2(\Gamma -1)}{1+2(\Gamma -1)}\,,  \label{eq45}
\end{equation}%
where $w_{B}$ is the state equation of the matter, in practice zero.
This equation is derived from a well known formula \cite{37}.

Now, the only two potentials for which $\Gamma $ is strictly constant are
the exponential ($\Gamma =1$) and the inverse power
\begin{equation}
V(\varphi )=M\varphi ^{-\alpha }\,,  \label{inversepot}
\end{equation}%
with $\alpha >0$ ($\Gamma =\left( \alpha +1\right) /\alpha $). In the first
case, from Eq. (\ref{eq45}) we get $w=w_{B}$, which seems to be untenable.
In the second case, we have
\begin{equation}
w=\frac{\alpha w_{B}-2}{\alpha +2}\,.  \label{trackw}
\end{equation}

These results have to be discussed and revised. So, let us first make the
following four considerations.

1) In our example, the asymptotic value of $w$ is $\ -1/2$, so that, also in
the case of practically constant $w$, it does not scale as the background.
In fact, it is not difficult to show that, with a generic exponential
potential $\ V=A\exp(-\lambda\varphi)$, a scaling equal to the matter
background is possible only for $\lambda>\sqrt{3}$. This seems to be in
contradiction with Eq. (\ref{eq45}), which makes no distinction on the value
of $\lambda$.

2) In the case of the double exponential of Eq. (\ref{doubpot}), it is easy
to show that it is $\Gamma<1$ always, and yet it has proved to be suitable
for tracking. This is not a contradiction with the tracking theorem, which
only claims on sufficiency of $\Gamma>1$; but, as a matter of fact, a bad
interpretation in the sense of necessity has driven the literature to
neglect an interesting field of study.

3) In the case of Eq. (\ref{inversepot}), the system should evolve
towards the value of Eq. (\ref{trackw}), but in Ref. \cite{38} they
show that this value is maintained only for a short period, while the
asymptotic value is $ w=-1$. It is in fact not difficult to show that,
if the scalar field dominates, and if there is a constant asymptotic
value for $w$, this must be $-1$.

4) If we insert our exact solution into the expression of $\Gamma $ in Ref.
\cite{37}, we do not find consistency! \medskip

We have thus worked out again the calculation for $\Gamma $, which is
detailed in the Appendix, and found that the result in Ref. \cite{37} is
incorrect. The correct result is indeed the following
\begin{equation}
\Gamma =1-\frac{2}{1+w}\frac{{\tilde{\tilde{x}}}}{\left( 6+\tilde{x}
\right)^{2}}-\frac{1-w}{2\left( 1+w\right) }\frac{\tilde{x}}{6+\tilde{x}}+3
\frac{w_{B}-w}{1+w}\frac{1-\Omega _{\varphi }}{6+\tilde{x}}\,,
\label{eqright}
\end{equation}
where
\begin{equation}
x\equiv \frac{1+w}{1-w}\quad ,\quad {\tilde{x}}\equiv \frac{d(\log
x)}{d(\log a)}\quad ,\quad {\tilde{\tilde{x}}}\ \equiv
\frac{d\,{\tilde{x}}}{d(\log a)}
\,,
\end{equation}
so that, in the case when $w$ is nearly constant, $\tilde{x}$ and
${\tilde{\tilde{x}}}$ can be neglected.

It is straightforward to check that this is consistent with our
solution, but what is really important is that, if $w$ is nearly
constant, Eq. (\ref {eq45}) is not true. In the case of $w$ nearly
constant, in fact, we have from the above equation that Eq.
(\ref{eq45}) must be substituted by
\begin{equation}
\Gamma \approx 1+\frac{w_{B}-w}{2(1+w)}(1-\Omega _{\varphi })\,,
\label{wright}
\end{equation}%
and we find the answers to points 1) and 3). Indeed, if $w$ scales as $w_{B}$
(and we saw that this can be the case only for great values of the parameter
$\lambda )$, then $\Omega _{\varphi }$ goes to a constant value $<1$, and
everything is consistent; but, if the scalar field dominates and go faster
forever, then $\Omega _{\varphi }\rightarrow 1$, so that $w$ can take any
value. It can be indeed computed as \cite{12}
\begin{equation}
w=\frac{\lambda ^{2}-3}{3}\,.
\end{equation}

We also see that, in the case of Eq. (\ref{inversepot}), since $
\Omega_{\varphi}\rightarrow1$, the only way to keep $\Gamma$ constant is to
let $w\rightarrow-1$. In this case, however, the last terms in Eq.
(\ref {eqright}) cannot be neglected.

We thus see that, now, the picture is consistent. But Eq.
(\ref{wright}) allows to obtain another interesting result. If there
is a period in which $ w $ is nearly constant and the scalar field
ultimately dominates, then the asymptotic value for $\Gamma$ is $1$
and the potential is \textit{obliged} to be exponential. However, we
must note that this only means that the true potential approximates an
exponential in the late evolution, without saying anything else on its
analytical expression. In fact, a strictly constant $w$ implies an
inverse hyperbolic sine potential \cite{3,39}.

\section{Conclusions}

We have presented above a class of physically meaningful exponential
potentials, which allow general exact solution of the Friedman and
Klein-Gordon equations. We have also shown that this solution, in the
limits of validity of the model, meets all the requests which are
generally made for a quintessence model, except one: the possibility
of switching off the acceleration in the future, which is needed for
the asymptotic freedom of the model \cite{28}. This can indeed be
obtained by a modification of the potential.

One possibility, studied in Refs. \cite{31,32,33}, is to bring the
value of $
\sigma$ to the range $\sqrt{2}<\sigma<\sqrt{3}$, so that the limit value for
$w$ is raised to a value greater than $-1/3$. Another possibility is
to pass to a combination of exponentials, which we have treated in
another paper \cite{23}.

This last example is particularly interesting, and it is worthwhile to
summarize some results. Using the potential
\begin{equation}
V(\varphi )=(Ae^{\sigma \varphi /2}-Be^{-\sigma \varphi /2})^{2}\,,
\label{doubpot}
\end{equation}%
it is possible to obtain essentially the same situation as shown here
(with a simple exponential potential) for the past, until present
epoch. Towards the future, instead, we get that $w$ oscillates forever
between $-1$ and $1$, but the oscillations of $\rho _{\varphi }$ turn
out to be damped, so that it effectively scales like matter, and dark
energy never reaches complete domination of dynamics. (See Ref.
\cite{23} for details, and Ref. \cite{33bis} for a closely related
situation.)

Coming back to the potential in Eq. (\ref{eq1}), it is possible to
guess a slightly positive spatial curvature. As said in the
introduction, this is not excluded by the present data. It is then
possible to add a curvature term into Eqs. (\ref{eq4}), (\ref{eq5})
and (\ref{eq6}), and try to integrate them numerically. An appropriate
choice for $\Omega_{k}$ gives the plot of Fig. 12, where we see that
the acceleration starts from negative values, reaches a regime of
positive values, and then goes back to negative in the future.

We have now to shortly discuss which are the shortcomings of the model
we have considered so far in this paper.

First of all, although exponential potentials are widely considered as
possible candidates to some kind of exotic dark energy (for physical
reasons which are \textit{a priori} with respect to the framework here
presented), we have found no explanation which justifies the
particular class of exponential potentials we have chosen, except the
nice feature of the Noether symmetry of the Lagrangian (\ref{eq7})
(but see Refs. \cite{26,27,29,30} for some discussion on this point).

Secondly, as we have repeatedly stressed, this model loses its
validity when radiation is not negligible, and we have seen that a lot
of interesting features occur just during that epoch. An extension is
thus needed, but this must probably be done with numerical methods.
Due to the very small values of some parameters, one must in fact be
very careful to avoid numerical errors. In order to generate properly
the plots of Sec. II, for instance, we had to work with $40$ digits
floating numbers, and there was not cumulated error typical of
numerical integration.

A study along these lines is presented in Ref. \cite{33}, where they find a
behavior very similar to the one depicted by us for a reasonable range of
values of $\sigma$ and initial conditions. They include radiation in
numerical computation, but the procedure is somewhat different by ours.

Most of all, our discussion shows not only that this model can
perfectly emulate a cosmological constant, but also gives no answers
about the real possibility to discriminate, at least between these two
models. It seems that we have shown that this discrimination is
however impossible on the pure basis of luminosity distance
observations. We have obtained this already known result \cite{34} by
showing that there is an important degeneracy in the $\Omega_{m}$
parameter.

In principle, this parameter can be measured independently, but it
appears very difficult to reduce its present error of a factor, say,
$10$ in the near future. We have also shown elsewhere that the two
models are equally compatible with the observations on the peculiar
velocities of galaxies \cite{35}. This kind of measure is also
affected by large errors, is time consuming and can say something only
at rather low redshift ($z\approx 0.15)$. More promising seems to be
the analysis of CMBR, as said above, and gravitational lensing,
particularly of very distant QSO (see Ref. \cite{36}, for instance,
and references therein). Here the problem appears to be the presence
of a lot of other parameters, leading to huge degeneracy.

In our opinion, however, the main result which we obtain in this paper
is probably pedagogical: we have learned that some statements, taken
for granted by the majority of people, can be misleading. Let us
summarize some of them here again.

We have seen that the most interesting epoch in our model is when $w$ is
mostly variable, but only with respect to $\log a$. This shows that to say
`` $w$ is necessarily almost constant'' is badly stated, and risks to be
merely a prejudice.

The study of the tracker behavior is subtle and strongly depends on the
representation chosen. This is so even in the case when a general exact
solution is given. This does not mean that qualitative analysis cannot be
done in less favourable situations, but that it should be very carefully and
unambiguously presented.

The requirement $\Gamma >1$ seems to be not necessary and, in any case, many
widely accepted statements on exponential potentials and tracking solutions
are indeed based on a result which has revealed to be incorrect.

In conclusion, we think that the final answer to the problem of the
acceptable form for the potential is far to be reached, and that more
insight on the physical motivations for the dark energy potential should be
developed.

\section*{Acknowledgments}

We are very grateful to A. Lewis (CITA, Univ. of Toronto and Astronomy
Dept., Harvard University) for invaluable help in running CosmoMC during his
visit to Naples and for suggestions and comments to the manuscript.

We are also grateful to M. Demianski (Dept. of Theoretical Physics, Univ. of
Warsaw) for reading and discussing the manuscript.

\appendix*
\section*{}

Let us rewrite Friedmann and Klein-Gordon equations

\begin{equation}
H^{2}=\frac{1}{3}\left( \rho_{B}+\rho_{\varphi}\right) \text{\quad },\text{
\quad}\dot{H}\equiv\frac{\ddot{a}}{a}-H^{2}=-\frac{1}{2}\left[
\rho_{B}\left( 1+w_{B}\right) +\rho_{\varphi}\left( 1+w\right) \right] \,,
\label{frid}
\end{equation}

\begin{equation}
\ddot{\varphi}+3H\dot{\varphi}+V^{\prime}=0 \,.
\end{equation}

We then define as above

\begin{equation}
x\equiv\frac{\dot{\varphi}^{2}}{2V}=\frac{1+w}{1-w}\quad,\quad\tilde{x}
\equiv \frac{d\ln x}{d\ln a}\ =\frac{\dot{x}}{H x}\quad,\quad{\tilde{
\tilde{x}}} \equiv\frac{d^{2}\ln x}{d\ln a^{2}}=\frac{1}{H}\frac{d}{dt}
\left( \frac{\dot{ x}}{H x}\right) \,.
\end{equation}

After substitution of

\begin{equation}
V^{\prime}\equiv\frac{dV}{d\varphi}=\frac{\dot{V}}{\dot{\varphi}}\quad,\quad
V^{\prime\prime}=\frac{dV^{\prime}}{d\varphi}
=\frac{1}{\dot{\varphi}}\frac{d }{dt}\left(
\frac{\dot{V}}{\dot{\varphi}}\right) =\frac{\ddot{V}}{\dot{
\varphi}^{2}} -\frac{\dot{V}\ddot{\varphi}}{\dot{\varphi}^{3}}
\end{equation}
into the definition of $\Gamma$, we get

\begin{equation}
\Gamma=\left( \frac{\ddot{V}}{\dot{\varphi}^{2}}-\frac{\dot{V}\ddot{\varphi}
}{\dot{\varphi}^{3}}\right) \frac{V\dot{\varphi}^{2}}{\dot{V}^{2}}=\frac{V
\ddot{V}}{\dot{V}^{2}}+\frac{V}{\dot{V}\dot{\varphi}}\left( 3H\dot{\varphi}
+V^{\prime}\right) =\frac{V\ddot{V}}{\dot{V}^{2}}+3H\frac {V}{\dot{V}}
+
\frac{V}{\dot{\varphi}^{2}}\, .
\end{equation}

Let us now compute the three terms of $\Gamma$. Starting from the last one,
we have

\begin{equation}
\frac{V}{\dot{\varphi}^{2}}=\frac{1}{2x}=\frac{1-w}{2\left( 1+w\right) } \,.
\end{equation}
As for the second term, let us observe that

\begin{eqnarray}
\frac{\dot{V}}{V} &=&\frac{d\ln V}{dt}=\frac{d}{dt}\left( {2\ln \dot{\varphi}
-\log 2-\log x } \right)  \nonumber\\
&=&\frac{2\ddot{\varphi}}{\dot{\varphi}}-\frac{\dot{x}}{x}
=\frac{2}{\dot{
\varphi}}\left( {-3H\dot{\varphi}-V^{\prime} }\right) -H{\tilde{x}}  \nonumber \\
&=&-6H-\frac{\dot{V}}{x V}-H{\tilde{x}}\,,
\end{eqnarray}
so that we have

\begin{equation}
3H\frac{V}{\dot{V}}=-3\frac{1+x}{\left( 6+{\tilde{x}}\right)
x}=-\frac{6}{1+w}
\frac{1}{6+{\tilde{x}}} \,.  \label{5.23}
\end{equation}

Finally, let us compute the first term

\begin{equation}
\frac{V\ddot{V}}{\dot{V}^{2}}=\frac{V}{\dot{V}}\frac{d\ln\dot{V}}{dt} \,.
\end{equation}
First, we have

\begin{equation}
\frac{d\ln\dot{V}}{dt}=\frac{\dot{V}}{V}+\frac{\dot{H}}{H}+\frac{\dot{x}}{x}
+ \frac{d{\tilde{x}}/dt}{6+{\tilde{x}}}-\frac{\dot{x}}{1+x} \,,
\end{equation}
and, being

\begin{equation}
\frac{d{\tilde{x}}}{dt}=\frac{d}{dt}\left( \frac{\dot{x}}{H x}\right) =H
{\tilde{\tilde{x}}} \,,
\end{equation}
we get

\begin{equation}
\frac{d\ln\dot{V}}{dt}=-\frac{6+{\tilde{x}}}{1+x}Hx+\frac{\dot{H}}{H} +H{\tilde{
x}}+\frac{H{\tilde{\tilde{x}}}}{6+{\tilde{x}}}-\frac{{\tilde{x}}}{1+x}Hx
\,.
\end{equation}

From this relation, together with Eq.(\ref{5.23}), we obtain

\begin{eqnarray}
\frac{V\ddot{V}}{\dot{V}^{2}} &=&1-\frac{\dot{H}}{H^{2}}\frac{1+x}{\left( 6+
{\tilde{x}}\right)
x}-\frac{1+x}{x}\frac{{\tilde{x}}}{6+{\tilde{x}}}-\frac{1+x}{x}
\frac{{\tilde{\tilde{x}}}}{\left( 6+{\tilde{x}}\right) ^{2}}+\frac{\tilde{x}
}{6+{\tilde{x}}}  \nonumber\\
&=&1-\frac{\dot{H}}{H^{2}}\frac{2}{1+w}\frac{1}{6+{\tilde{x}}}-\frac{1-w}{1+w}
\frac{{\tilde{x}}}{6+{\tilde{x}}}-\frac{2}{1+w}\frac{{\tilde{\tilde{x}}}}{
\left( 6+{\tilde{x}}\right) ^{2}}\,,
\end{eqnarray}
and, considering that

\begin{equation}
\frac{2x}{1+x}=1+w \,,
\end{equation}
we find

\begin{equation}
\Gamma =1-\frac{2}{1+w}\frac{{\tilde{\tilde{x}}}}
{\left( 6+\tilde {x}
\right) ^{2}}-\frac{1-w}{1+w}\frac{\tilde{x}}{6+\tilde{x}}-\frac{6}{1+w}
\frac{1}{6+\tilde{x}} \\
-\frac{\dot{H}}{H^{2}}\frac{2}{1+w}\frac{1}{6+\tilde{x}}+\frac{1-w}{ 2\left(
1+w\right) } \,.
\end{equation}

We have now to eliminate the term $\dot{H}/H^{2}$. From Eq. (\ref{frid}) it
is

\begin{eqnarray}
\frac{\dot{H}}{H^{2}} &=&-\frac{3}{2}\frac{\rho _{B}\left( 1+w_{B}\right)
+\rho _{\varphi }\left( 1+w\right) }{\rho _{B}+\rho _{\varphi }}=-\frac{3}{2}
\left( 1+\frac{\rho _{B}w_{B}+\rho _{\varphi }w}{\rho _{B}+\rho _{\varphi }}
\right)  \nonumber \\
&=&-\frac{3}{2}\left[ 1+\left( 1-\Omega _{\varphi }\right) w_{B}+\Omega
_{\varphi }w\right] =\frac{3}{2}\left[ \Omega _{\varphi }\left(
w_{B}-w\right) -\left( 1+w_{B}\right) \right] \,,
\end{eqnarray}
and, by substitution, we eventually get

\begin{equation}
\Gamma =1-\frac{2}{1+w}\frac{\tilde{\tilde{x}}}{\left( 6+\tilde{x}
\right) ^{2}}-\frac{1-w}{2\left( 1+w\right) }\frac{\tilde{x}}{6+\tilde{x}}+3
\frac{w_{B}-w}{1+w}\frac{1-\Omega _{\varphi }}{6+\tilde{x}}\,,
\end{equation}
which should be compared with the formula reported in Ref. \cite{37}

\begin{equation}
\Gamma =1+\frac{w_{B}-w}{2(1+w)}-\frac{1+w_{B}-2w}{2\left( 1+w\right) }\frac{
\tilde{x}}{6+\tilde{x}}-\frac{2}{1+w}\frac{\tilde{\tilde{x}}}{\left( 6+
\tilde{x}\right) ^{2}}\,.
\end{equation}

\begin{table}[tbp] \centering

\begin{tabular}{|c|c|c|c|c|c|c|}
\hline
& \multicolumn{3}{|c|}{$\Lambda $-term $-\log (LH)=765.3$} &
\multicolumn{3}{|c|}{Exp. pot. $-\log (LH)=767.3$} \\ \hline
par. & best fit & lower & upper & best fit & lower & upper \\ \hline
$\Omega _{b}h^{2}$ & $0.0226$ & $0.0206$ & $0.0256$ & $0.023$ &
$0.0213$ & $ 0.0266$ \\ \hline $\Omega _{dm}h^{2}$ & $0.120$ & $0.103$
& $0.139$ & $0.110$ & $0.094$ & $ 0.134$ \\ \hline $n_{s}$ & $0.960$ &
$0.914$ & $1.05$ & $0.948$ & $0.905$ & $1.04$ \\ \hline $\Omega
_{m}$ & $0.298$ & $0.222$ & $0.379$ & $0.298$ & $0.232$ & $0.383$ \\
\hline
$z_{re}$ & $12.1$ & $2.57$ & $24.0$ & $12.6$ & $2.50$ & $23.6$ \\ \hline
$h$ & $0.692$ & $0.643$ & $0.770$ & $0.669$ & $0.628$ & $0.729$ \\ \hline
\end{tabular}
\caption{Best fit parameters for the two models. LH is the likelihood
function obtained with the CosmoMC codes, as explained in the text.
Caution must be paid in the interpretation of the lower and upper
limits of the parameter values. They are not the $1\sigma$
constraints, but the extremal points of the $6$-dimensional confidence
region. In any case, this table gives a clear indication that the two
models are compatible with usual data. The small difference in the
likelihood is not significant.}

%\QTP{ymb}

\end{table}

\begin{figure}[tbp]
\includegraphics{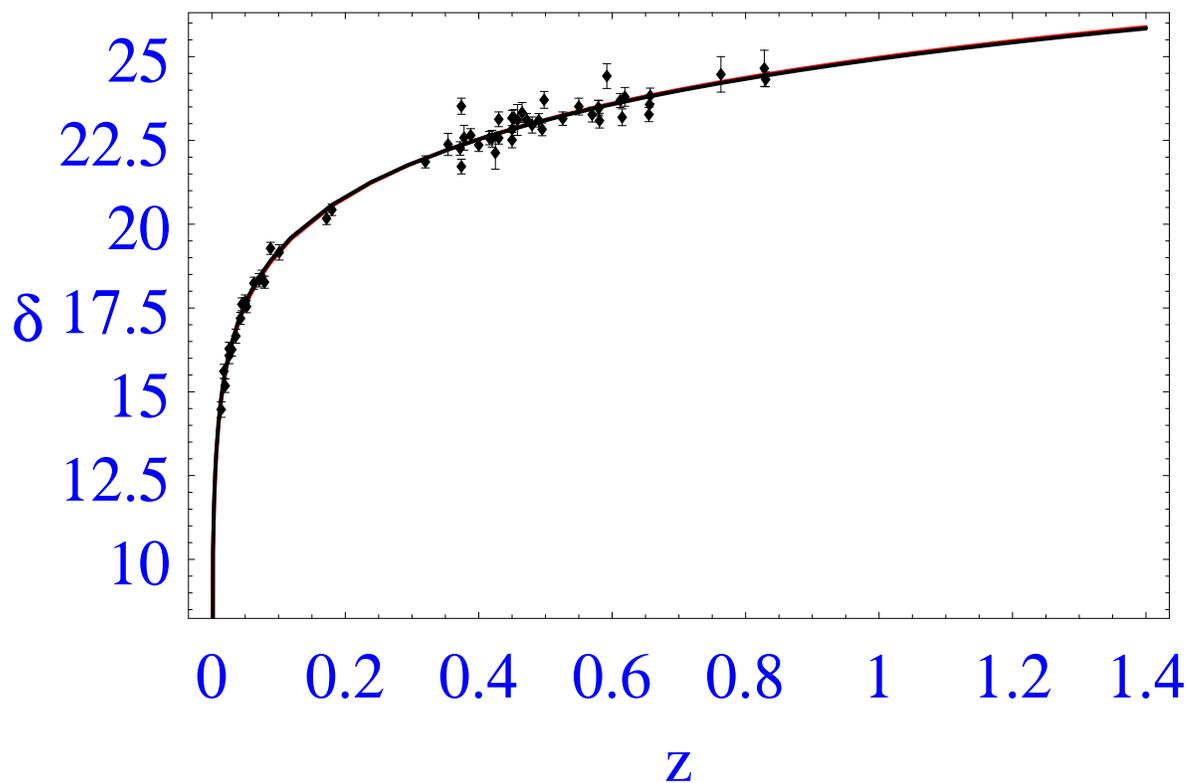}
\caption{The distance modulus for the $\Lambda$-term and the exponential
potential models, obtained from best fit parameters of Table I, is compared
with SNIa Perlmutter at al. data. The two curves practically coincide.}
\end{figure}
%\FRAME{ftbp xF }{}{}{}{}{}{}{

\begin{figure}[tbp]
\includegraphics{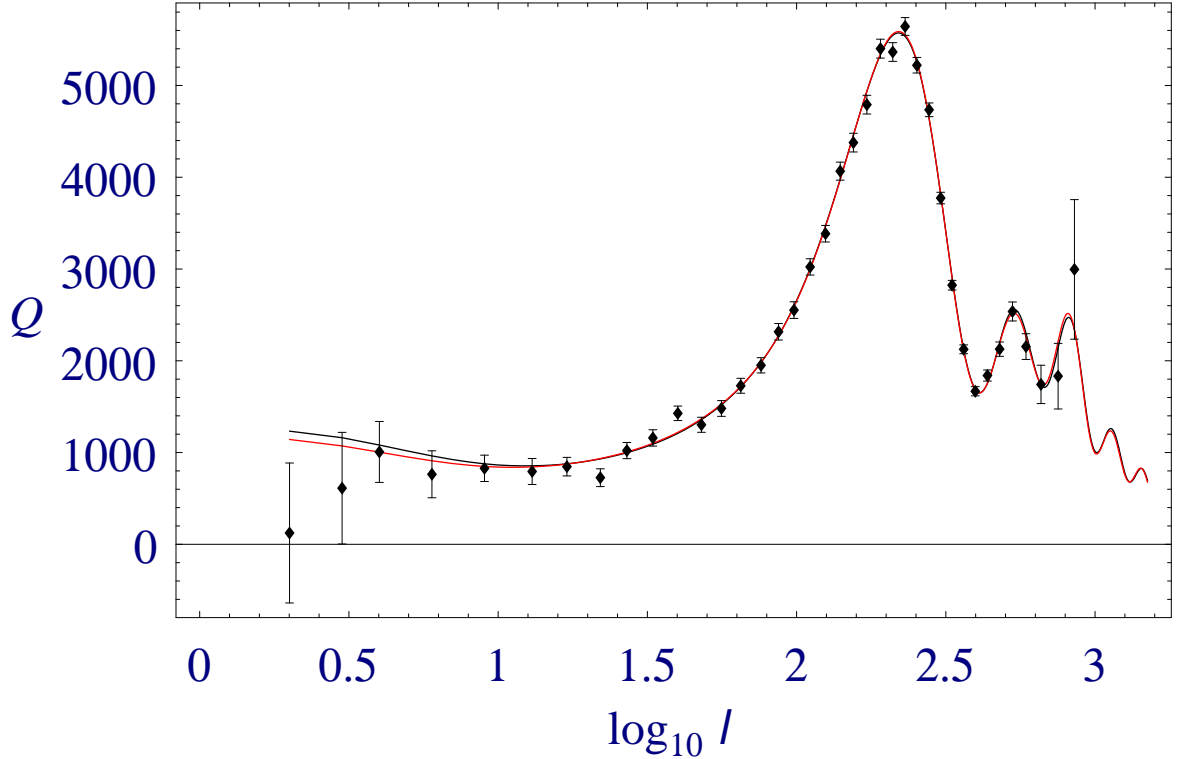}
\caption{The CMBR angular power spectrum $Q \equiv l(l+1)C_l/2\protect\pi$
for the two models, obtained with CAMB codes from the best fit
parameters of Table I. The two curves practically coincide except but
for small $l$'s, where the exponential potential gives slightly higher
values.}
\end{figure}
%\FRAME{ftbp xF }{}{}{}{}{}{}{

\begin{figure}[tbp]
\includegraphics{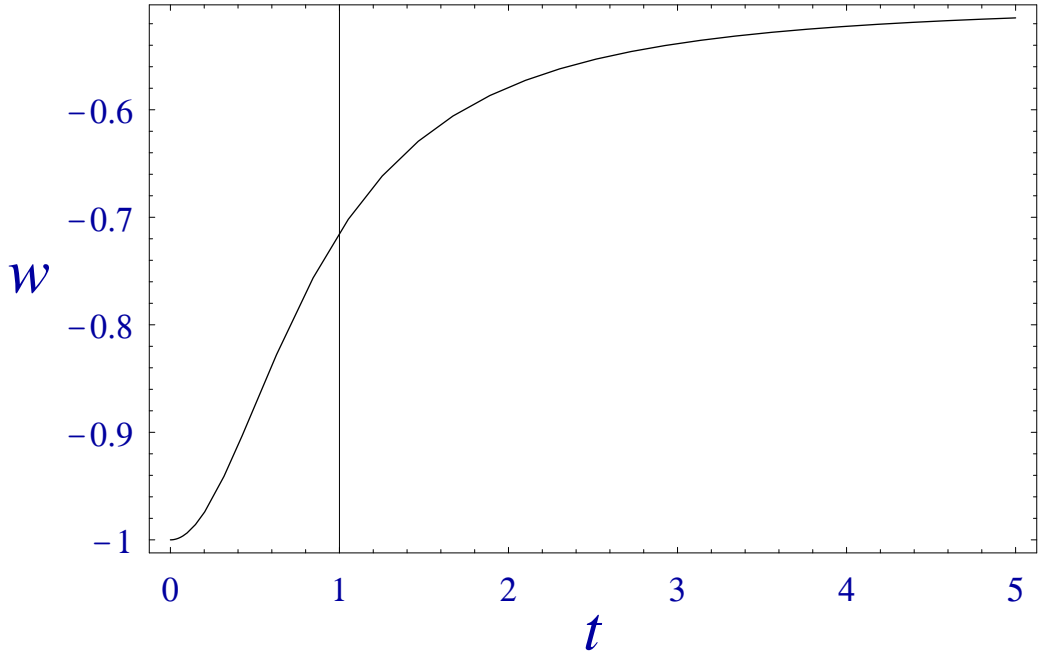}
\caption{Plot of the scalar-field equation of state versus time. The
vertical bar marks present time. An almost constant behavior is
obtained only in the limit of large t.}
\end{figure}
%\FRAME{ftbp xF }{}{}{}{}{}{}{

\begin{figure}[tbp]
\includegraphics{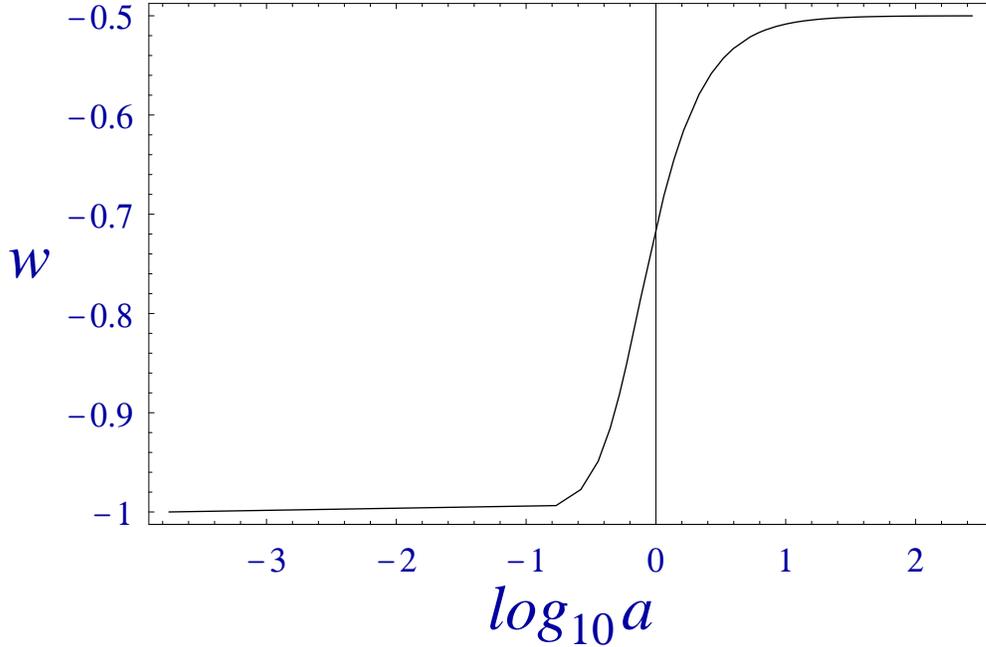}
\caption{Plot of the scalar-field equation of state versus ${\log}_{10}a$
with the best fit value of $\Omega_m = 0.298$. The vertical bar marks
${\log}_{10} a_0$. Only with this choice of variables, there is
evidence of a transition from $w \approx -1$ in the past to
$w\approx-0.5$ in the future. (See Fig. 6 for further comments.)}
\end{figure}
%\FRAME{ftbp xF }{}{}{}{}{}{}{

\begin{figure}[tbp]
\includegraphics{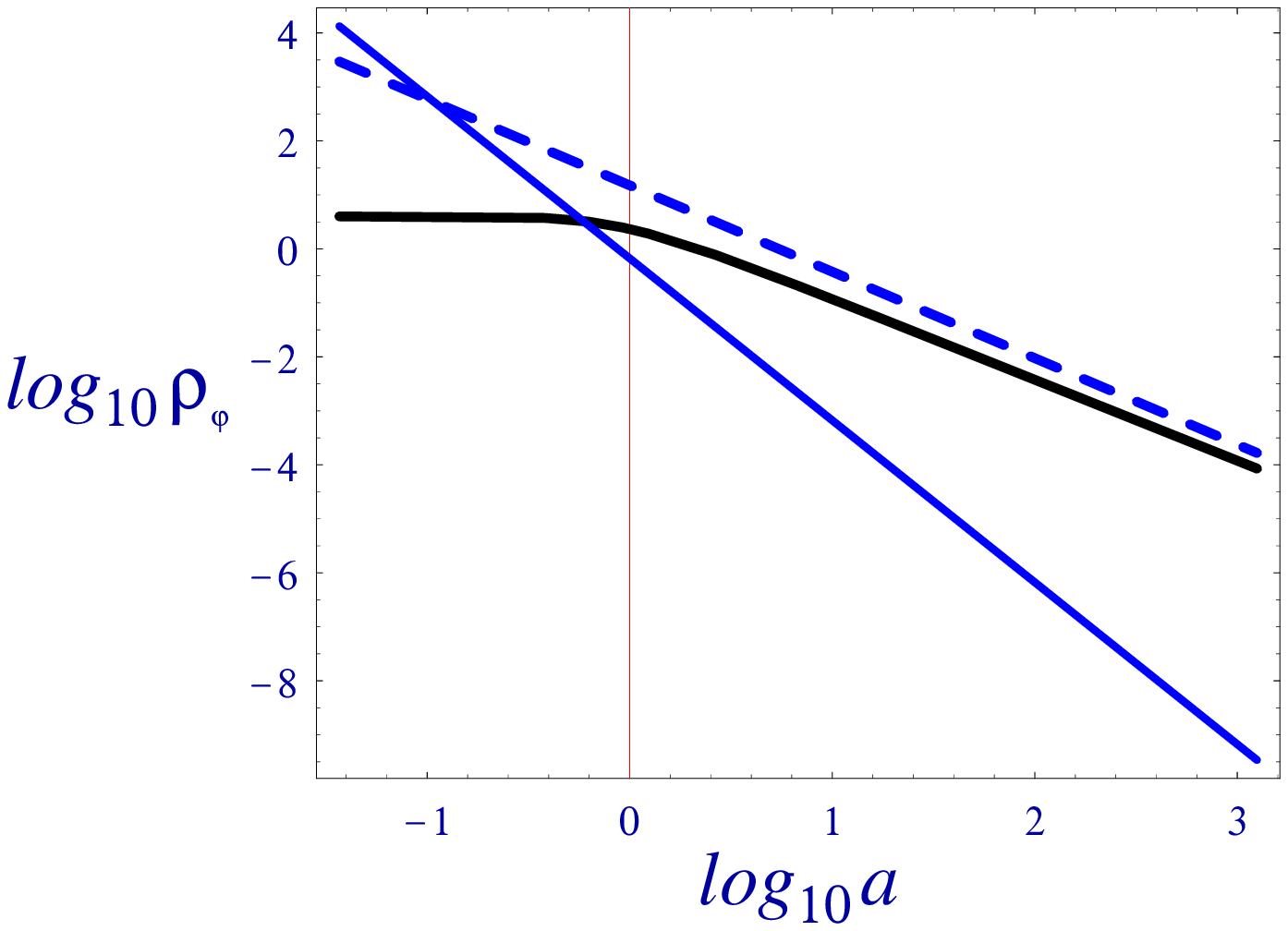}
\caption{Plot of ${\log}_{10} \protect\rho_\protect\varphi$ versus ${\log}_{10} a$. The vertical bar again marks ${\log}_{10}a_0$. The full straight
line indicates the log-log plot of $\protect\rho_m$ versus a, while
the dashed one stays just for reference of the the asymptotic
behavior.}
\end{figure}
%\FRAME{ ftbp xF }{}{}{}{}{}{}{

\begin{figure}[tbp]
\includegraphics{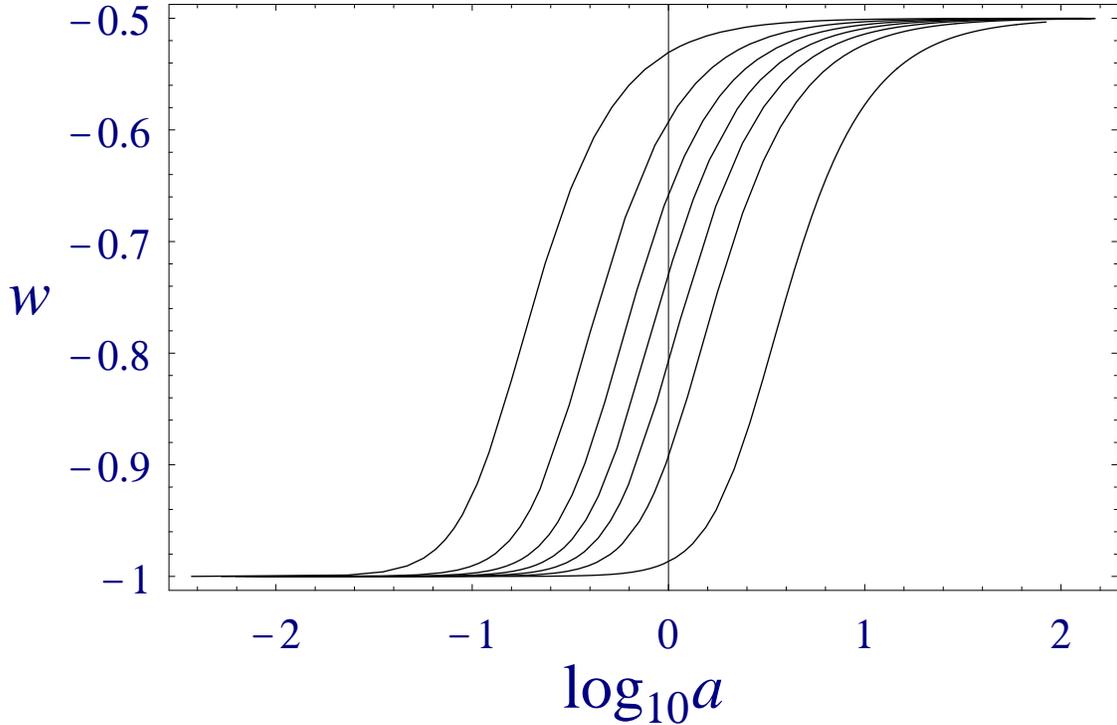}
\caption{Plot of the scalar-field equation of state versus ${\log}_{10} a$
for various values of $\Omega_m$ in the range $0.1$ to $0.9$. The
vertical bar marks ${\log}_{10}a_0$. We see, as better explained in
the text, that the situation is qualitatively similar for any
reasonable choice of $\Omega_m$.}
\end{figure}
%\FRAME{ ftbp xF }{}{}{}{}{}{}{

\begin{figure}[tbp]
\includegraphics{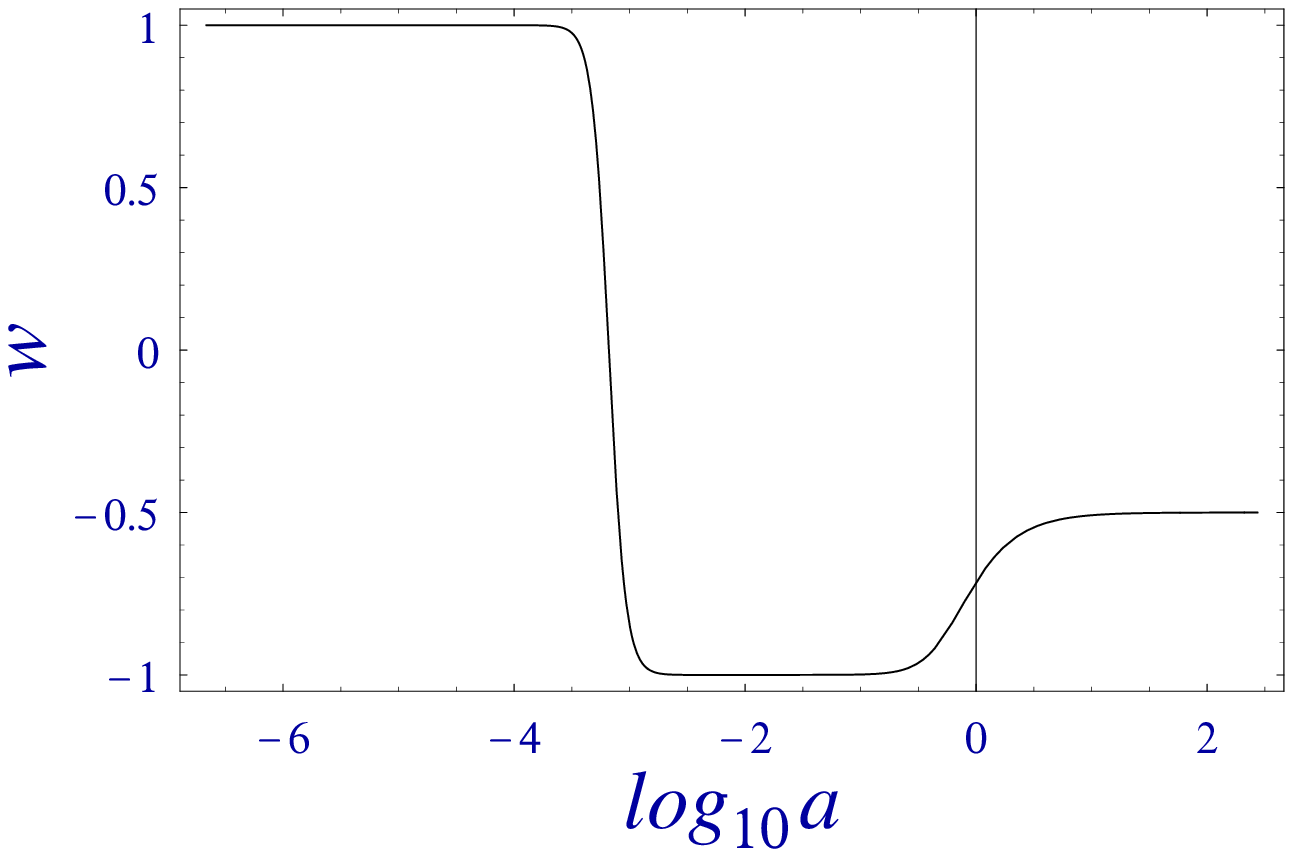}
\caption{Plot of the scalar-field equation of state versus ${\log}_{10} a$
after releasing the assumption $u_2 = v_2 = 0$. As usual, The vertical
bar marks ${\log}_{10}a_0$.}
\end{figure}
%\FRAME{ftbp xF }{}{}{}{}{}{}{

\begin{figure}[tbp]
\includegraphics{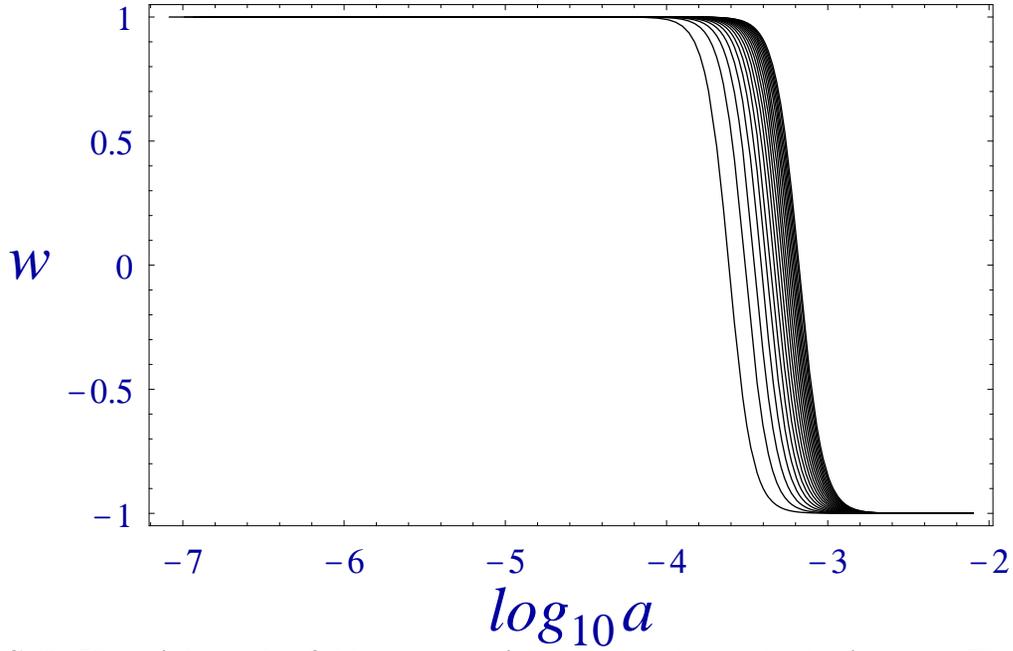}
\caption{Plot of the scalar-field equation of state versus ${\log}_{10} a$
in the far past. The first \textquotedblleft phase transition
\textquotedblright is shown for various values of $\protect\varepsilon
\equiv u_1 v_2$. Only a limited range of such values is here used. The full
range would imply the left part of the plot completely filled in with
curves.}
\end{figure}
%\FRAME{ftbp xF }{}{}{}{}{}{}{

\begin{figure}[tbp]
\includegraphics{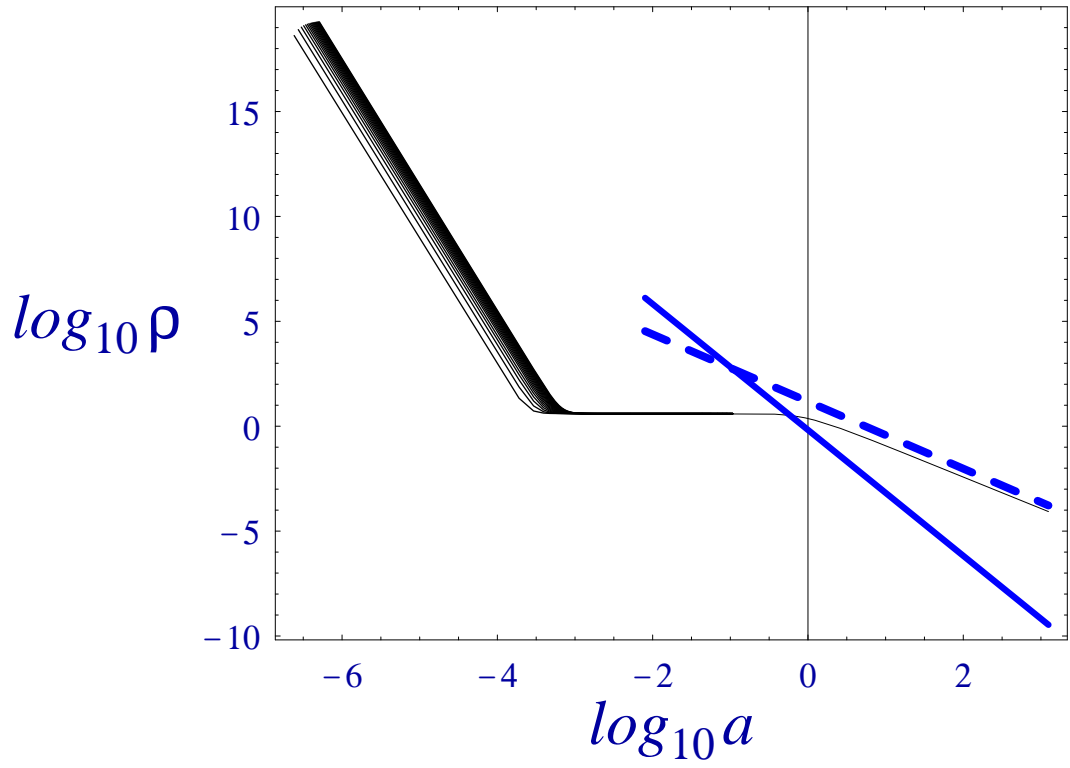}
\caption{The same as in Fig. 5, but with $\protect\varepsilon \neq 0$.
Again, the left part should be imagined as filled in with curves.}
\end{figure}
%\FRAME{ ftbp xF }{}{}{}{}{}{}{

\begin{figure}[tbp]
\includegraphics{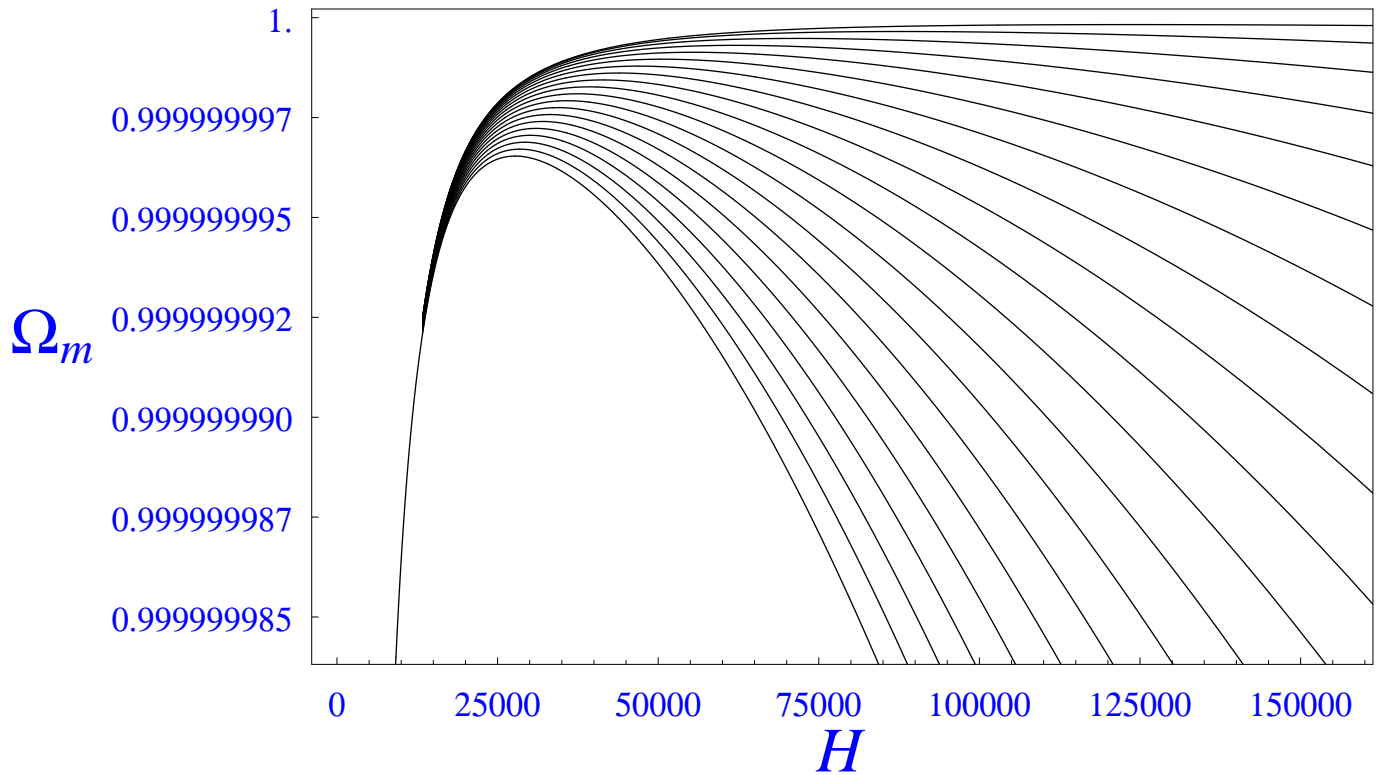}
\caption{Plot of $\Omega_m$ versus $H$, with the same assumptions as in
Figs. 8 and 9. We again see a tracking behavior, but with a more
limited range of values. In particular, the variation of $\Omega_m$ is
practically irrelevant.}
\end{figure}
%\FRAME{ftbp xF }{}{}{}{}{}{}{

\begin{figure}[tbp]
\includegraphics{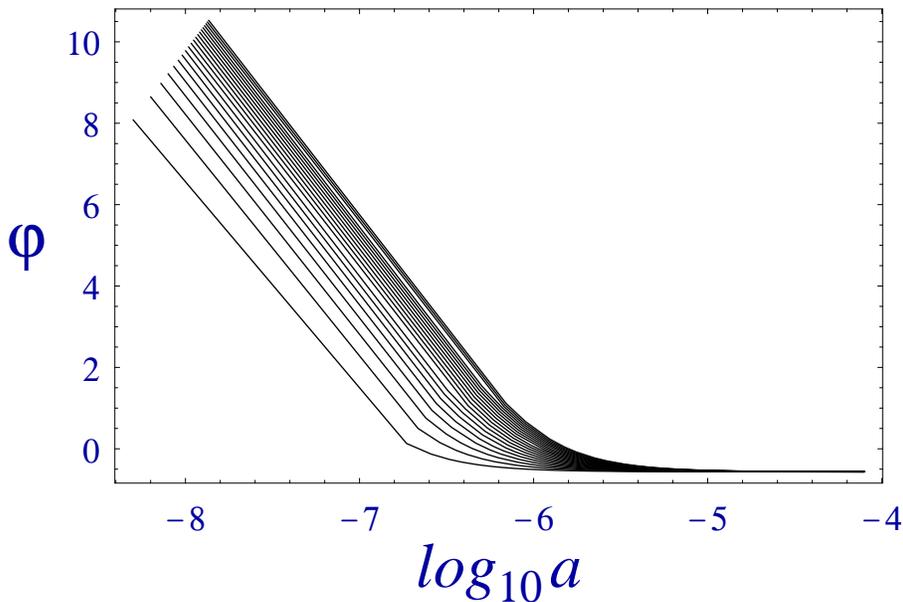}
\caption{Tracking behavior is obtained for $\protect\varphi$, only if we
arbitrarily assume a fixed value for $\protect\varphi_0$ (in this
case, $\protect\varphi_0 = 0$).}
\end{figure}
%\FRAME{ftbp xF }{}{}{}{}{}{}{

\begin{figure}[tbp]
\includegraphics{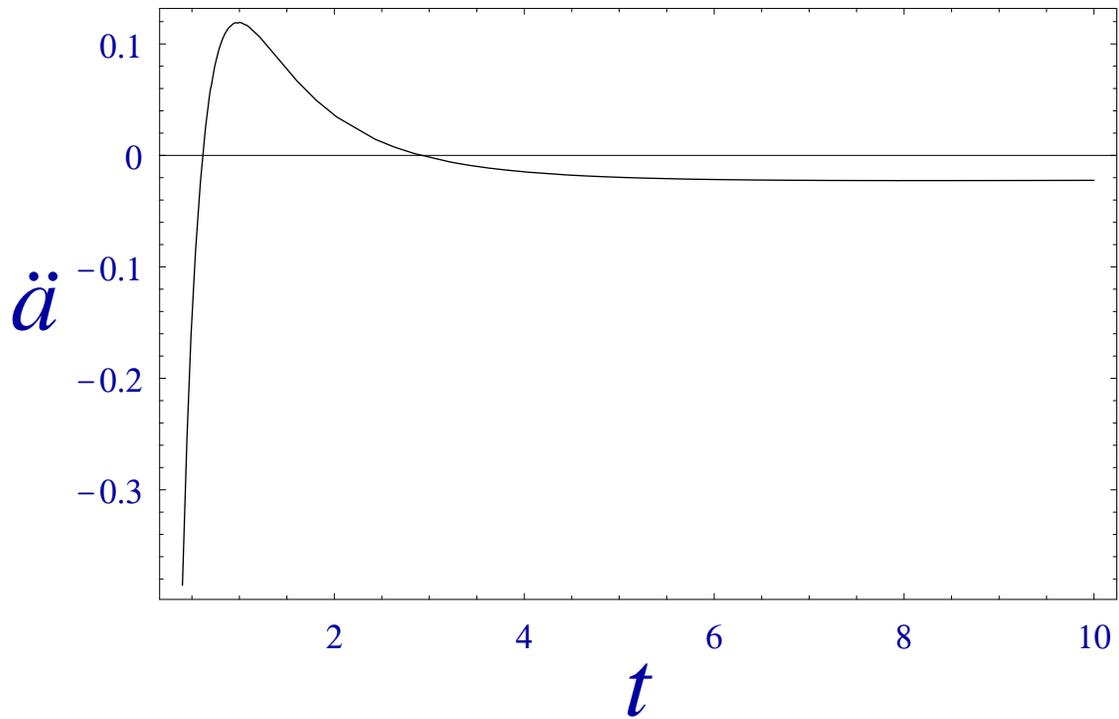}
\caption{Plot of the acceleration with respect to time, obtained from
numerical integration in the case $\Omega_k > 0$. The units are
arbitrary, and this figure only aims to show that the problem of
eternal acceleration could be avoided still maintaining $\Omega_k$ in
the range allowed by present constraints.}
\end{figure}

\end{document}